\documentclass[aps,prl,twocolumn,amsmath,amssymb,showpacs,superscriptaddress,notitlepage,longbibliography,10pt]{revtex4-1}

\usepackage[colorlinks=true,linkcolor=blue,anchorcolor=red,citecolor=blue, urlcolor=blue]{hyperref}
\usepackage{bm}
\usepackage{graphicx}
\usepackage{color}
\usepackage{multirow}
\usepackage{extarrows}
\usepackage{cases}
\begin{document}
%\begin{CJK}{GBK}{kai}

\title{Electrical switching of altermagnetism}

\author{Yiyuan Chen$^\dag$}
\affiliation{State Key Laboratory of Quantum Functional Materials, Department of Physics, and Guangdong Basic Research Center of Excellence for Quantum Science, Southern University of Science and Technology (SUSTech), Shenzhen 518055, China}

\author{Xiaoxiong Liu$^\dag$}
\affiliation{State Key Laboratory of Quantum Functional Materials, Department of Physics, and Guangdong Basic Research Center of Excellence for Quantum Science, Southern University of Science and Technology (SUSTech), Shenzhen 518055, China}

\author{Hai-Zhou Lu}
\email{Corresponding author: luhz@sustech.edu.cn}
\affiliation{State Key Laboratory of Quantum Functional Materials, Department of Physics, and Guangdong Basic Research Center of Excellence for Quantum Science, Southern University of Science and Technology (SUSTech), Shenzhen 518055, China}
\affiliation{Quantum Science Center of Guangdong-Hong Kong-Macao Greater Bay Area (Guangdong), Shenzhen 518045, China}

\author{X. C. Xie}
\affiliation{International Center for Quantum Materials, School of Physics, Peking University, Beijing 100871, China}
\affiliation{Institute for Nanoelectronic Devices and Quantum Computing, Fudan University, Shanghai 200433, China}
\affiliation{Hefei National Laboratory, Hefei 230088, China}

\begin{abstract}
Switching magnetism using only electricity is of great significance for industrial applications but remains challenging. We find that, altermagnetism, as a newly discovered unconventional magnetism, may open an avenue along this effort. Specifically, to have deterministic switching, i.e., reversing current direction must reverse magnetic structure, parity symmetry has to be broken. We discover that, due to their symmetry which depends on chemical environments, altermagnet devices may naturally carry
the parity symmetry breaking required for deterministic electrical switching of magnetism. More importantly, we identify MnTe bilayers (Te-Mn-Te-Mn-Te) as candidate devices, with the help of symmetry analysis, first-principles calculations, and magnetic dynamics simulations. This scheme will inspire further explorations on unconventional magnetism.
\end{abstract}

\maketitle

\begin{figure}[tp]
\includegraphics[width=0.48\textwidth]{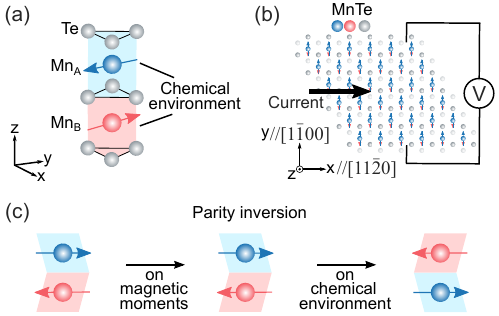}
\caption{(a) Crystal and magnetic structures of an altermagnet candidate MnTe. Te atoms around Mn atoms form different chemical environments (cyan and pink shadows). (b) Schematic of the setup for deterministic switching of altermagnetism by a pure electric current (black arrow) along the [11$\bar{2}$0] direction. The N\'eel order can be read out through an anomalous Hall voltage $V$ measured in the [$1\bar{1}$00] direction. (c) Parity symmetry has to be broken in any schemes of all-electric deterministic switching of magnetism.
When applying parity inversion to a MnTe bilayer, the magnetic moments (blue and red arrows) do not change, but the inversion of the chemical environments leads to the broken parity symmetry required for all-electric deterministic switching of altermagnetism. The bilayer here is more about the two Mn layers that solely contribute to the magnetic dynamics, although three Te layers are needed to maintain the altermagnetism. The effect of substrate can be balanced by capping the device.}
\label{Fig:MnTe}
\end{figure}

{\color{blue}\emph{Introduction}} -
Magnetic materials have been a focal point in condensed matter
physics and materials science, brewing diverse applications in technology,
energy storage, and sensing. While ferromagnetic materials are
well applied, antiferromagnetic materials offer advantages like quicker
dynamics and resilience to external magnetic fields \cite{Wadley2016Scn,Priyanka2020Scn,Park2011NatMat,Wang2012PRL,Marti2014NatMat}.
Recently, there has been
increasing interest in altermagnetism \cite{Wu04PRL,Wu07PRB,Hayami19jpsj,yuanLD20PRB,Hayami20prb,LiuJW21NC,Libor22PRXLandscape,Gu25PRL,Duan25PRL}, a new category of collinear antiferromagnetic phase, offering unique
properties and potential applications \cite{Gu25PRL,Duan25PRL,Lin25prl,Zhou25nat}.
The symmetry of altermagnets is determined
not only by the magnetic moments but also the chemical environments (Fig.~\ref{Fig:MnTe}), dictated by spin space groups \cite{Liu22prx,Libor22PRX2}. Their rotation-related sublattices with opposite magnetic moments breaks time reversal symmetry, holding the potential to exhibit other ferromagnetic features such as an anomalous Hall effect  \cite{Libor20SciAdv,Rafael21PRL,Naka19NC,Bose22NE,Osumi24PRB,Krempask24Nat,Reimers24NC,Naka20PRB,Lee24PRL}, which offers a better readout signal in information storage, compared to the previous antiferromagnets.
%{\color{red}Recently a new trend of tuning the band polarization with ferroelectricity has risen \cite{Gu25PRL,Duan25PRL}.}
On the other hand, using pure electrical methods to switch magnetism is a long-standing goal for facilitating applications, but switching altermagnetism still relies on magnetic fields \cite{Han2024SciAdv}, which is highly unfavorable. However, the mechanism of pure electrical switching of altermagnetism remains an important and open question.

In this Letter, we propose a scheme to switch altermagnetism deterministically using a pure electric current. We discover that, due to their symmetry which depends on chemical
environments, altermagnet devices may naturally carry the parity symmetry breaking required for
deterministic electrical switching of magnetism.
We identify MnTe and FeS bilayers (Te-Mn-Te-Mn-Te and S-Fe-S-Fe-S) as candidate devices, based on a search of many promising material candidates, with the help of rigorous symmetry analysis, solid first-principles calculations, and magnetic dynamics simulations. More importantly, our scheme can be generalized to other altermagnets, to inspire more future explorations.

\begin{figure*}[htpb]
\includegraphics[width=0.95\textwidth]{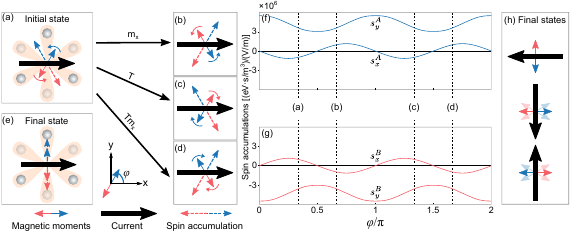}
\caption{Deterministic switching of altermagnetism in a MnTe bilayer by a pure electric current. (a) Assume an electric current (big black arrow) in the $+x$ direction and an initial N\'eel order (measured by the angle $\varphi$) of Mn magnetic moments (blue and red solid arrows). The light yellow shadow indicates six stable positions of the N\'eel order with a minimal energy, which happen to be aligned with the Te atoms (gray balls). The current can induce $\varphi$-dependent spin accumulations (with polarizations indicated by the dashed arrows) on the Mn atoms. The magnetic moments tend to chase the polarizations of spin accumulations (with the tendency indicated by the curved arrows), resulting the final N\'eel order in (e).
[(b)-(d)] Other possible initial N\'eel orders all have the same tendency towards the final N\'eel order in (e), because they can be related to the initial N\'eel order in (a) by the symmetry operations $m_x$, $T$, and $Tm_x$, where $m_x$ is the mirror reflection with
respect to the $y$-$z$ plane and $T$ is time reversal. Therefore, the $+x$-direction current can deterministically switch the N\'eel order to the $+y$ direction in (e). [(f) and (g)] The spin accumulations as functions of $\varphi$, calculated by using the Kubo formula and first-principles calculations  [see from Eqs.~(\ref{Eq:Kubo-d}) through (\ref{Eq:rho})]. $s^A_x$ means the $x$-direction component of spin accumulations on Mn atom A, and so on. The dashed lines mark the initial N\'eel order in panels (a), (b), (c), and (d), respectively. (h) The final N\'eel orders in the MnTe bilayer tend to be perpendicular to the driving current, as shown for currents in $-x$ and $\pm y$ directions.  After turning off the current, the N\'eel order tends to relax to the nearby stable positions, as shown by the light-color solid arrows.}
\label{Fig:Symmetry}
\end{figure*}

{\color{blue}\emph{Symmetry analysis of deterministic switching}} - Parity symmetry breaking is always required for a magnet to be deterministically switched by a pure electric current, i.e., the magnetic order has to be reversed when the electric current is reversed.
Reversing the current, which is a radial vector, corresponds to applying a parity inversion, under which however a magnetic order remains invariant if the magnet has parity symmetry. In other words, if the magnet has parity symmetry, opposite currents may yield the same magnetic order, which is against the deterministic requirement.
%Therefore, we need to break parity symmetry, to realize the all-electric deterministic switching of altermagnetism.

Parity symmetry can be broken in altermagnet devices due to their symmetry that depends on chemical environments. We will focus on a promising candidate of altermagnet, MnTe, though our scheme can be generalized to more altermagnets.
MnTe has a large spin splitting ($\sim$370 meV), high Curie temperature ($\sim$267 K) \cite{Lee24PRL}, and clear anomalous Hall signal \cite{Gonzalez2023PRL}.
Te atoms in MnTe form octahedra around Mn atoms, creating different chemical environments around the A, B Mn atoms, as shown by the cyan and pink shadows in Fig. \ref{Fig:MnTe}(a). The bulk crystal of MnTe has parity symmetry due to the periodic boundary condition in the [0001] crystallographic direction (defined as the $z$ direction here). To break parity symmetry for deterministic switching as well as to facilitate device applications, we consider a bilayer structure grown along the $z$ direction.

More importantly, as shown in Fig.~\ref{Fig:MnTe}(c), if the parity inversion is applied only to the magnetic moments of Mn atoms, the magnetic structure remains invariant, but the inversion of the chemical environments gives rise to the broken parity symmetry required for deterministic switching. This is a significant property, showing the advantage and potential of the MnTe bilayer as well as other altermagnets in pure electrical switching.

\begin{figure*}[htbp]
\includegraphics[width=0.85\textwidth]{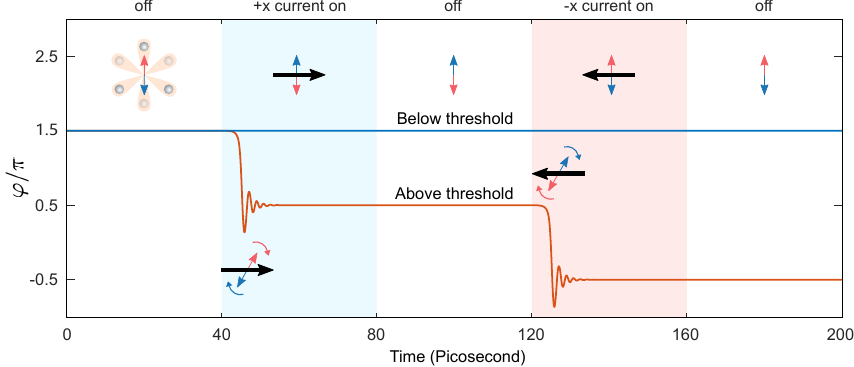}
\caption{Simulation of the switching dynamics in the MnTe bilayer. When applying a pure electric current (black arrow) above the threshold current density $j_c$, the N\'eel order between $\pm y$ directions ($\varphi= \pm \pi/2$, $3\pi/2=-\pi/2$) can be deterministically switched. The insets show the switching tendency (curved arrows) and status (red and blue solid arrows) of the N\'eel order. Here $j_c\approx2.5\times10^5$ A/cm$^2$ for the parameters $J$ = -1.756 eV, $\lambda$ = 0.01 meV, $K$ = 0.1 meV, and $J_m$ = 4.46 meV.}\label{Fig:Simulation}
\end{figure*}

{\color{blue}\emph{Deterministic switching driven by current-induced spin accumulations}} - Microscopically, the deterministic switching is guaranteed by
the symmetry of the spin accumulations induced by the current. The spin accumulations exert torques to rotate the magnetic moments of the Mn atom.
Deterministic switching means that no matter the initial N\'eel order, there must be a one-to-one correspondence between the direction of electric current and final N\'eel order (defined by $\mathbf{m}_A-\mathbf{m}_B$, where $\mathbf{m}_A$ and $\mathbf{m}_B$ are the magnetic moments on the  Mn atoms, as indicated by the blue and red arrows in Fig.~\ref{Fig:Symmetry}). In addition, the final N\'eel order has to be aligned with its corresponding spin accumulations, so that the torque vanishes and the rotation of the Mn moments stops.

MnTe has six stable positions for the N\'eel order, where the energy is minimal \cite{Kriegner17prb}, as indicated by the shadows in Fig.~\ref{Fig:Symmetry}(a). Without loss of generality, we assume a current in the $+x$ direction and a N\'eel order at an angle $\varphi$ measured from the $x$ direction. It is known that the Mn moments in MnTe are always polarized in the $x$-$y$ plane \cite{Gonzalez2023PRL}, so the two Mn moments in the bilayer can always be transformed to each other by a mirror reflection $m_z$ with respect to the $x$-$y$ plane so the N\'eel order remains invariant, while the two Mn moments always have opposite current-induced spin accumulations,
\begin{eqnarray}
m_z &: & \varphi \rightarrow \varphi, \nonumber\\
&& s^{A,B}_x(\varphi)=-s^{B,A}_x(\varphi), \nonumber\\ &&s^{A,B}_y(\varphi)=-s^{B,A}_y(\varphi),
\end{eqnarray}
which guarantees the same rotation tendency (as indicated by the curved arrows in Fig.~\ref{Fig:Symmetry}) for the two Mn moments.

We numerically calculate the spin accumulations as functions of $\varphi$ that describes the N\'eel order, by using the Kubo formula and first-principles calculations (see \hyperlink{Appendix A}{Appendix A}).
%[see from Eqs.~(\ref{Eq:Kubo-d}) through (\ref{Eq:rho})],
Figures.~\ref{Fig:Symmetry}(f) and \ref{Fig:Symmetry}(g) show the $x$ and $y$ components of the spin accumulations on the A and B Mn atoms, driven by the $+x$-direction current. In the present case, the moments are polarized in the $x$-$y$ plane because of the magnetic anisotropy, so we only need to consider the spin accumulations polarized in the $x$-$y$ plane \cite{Gonzalez2023PRL}. As shown in Fig.~\ref{Fig:Symmetry}(a), for the initial N\'eel order (solid arrows), the calculated spin accumulations are polarized along the dashed arrows, which leads to a rotation tendency towards the $y$ direction as indicated by the curved arrows. Roughly speaking, the moments $\mathbf{m}_a$ tend to chase the polarization of the spin accumulations $\mathbf{s}_a$, until $\mathbf{m}_a$ is aligned with $\mathbf{s}_a$ to cease the rotation, approximately governed by the spin-orbit torque \cite{Haney2008JMMM,Manchon2019handbook,Manchon2019RMP}
\begin{eqnarray}
    \mathbf{T}_a\sim \mathbf{m}_a\times \mathbf{s}_a.
\end{eqnarray}
Our calculated spin accumulations in Figs.~\ref{Fig:Symmetry}(f) and \ref{Fig:Symmetry}(g) show that, for arbitrary $\varphi$ in the first quadrant ($\varphi\in[0,\pi/2]$),
the $+x$-direction current always switch the N\'eel order anticlockwisely to the $+y$ direction.

More importantly, all possible initial N\'eel orders can be classified into the four quadrants in Figs.~\ref{Fig:Symmetry}(a)-\ref{Fig:Symmetry}(d), which all show the same rotation tendency towards the $+y$ direction for the $+x$-direction current, because
the initial N\'eel order in Fig.~\ref{Fig:Symmetry}(a) can be related to those in the other three quadrants by three different symmetry operations that transform the N\'eel order while keeping the crystal structure unchanged and Mn atoms A and B unexchanged (see details in \hyperlink{Appendix B}{Appendix B}).
Therefore, we conclude that the electric current in the $+x$ direction will always deterministically switch the N\'eel order to the $+y$ direction.

\begin{figure*}[tp]
\includegraphics[width=0.9\textwidth]{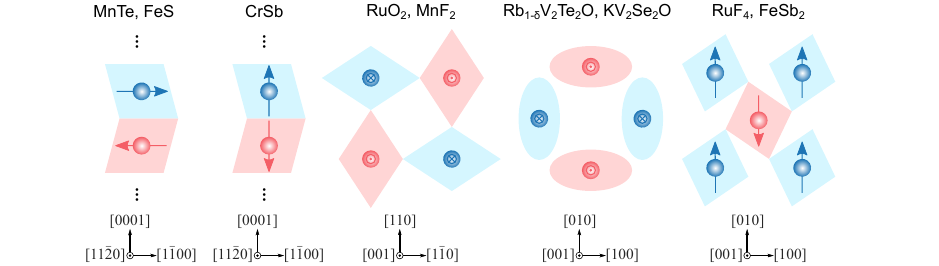}
\caption{Symmetry analysis of whether the material candidates for altermagnets can be deterministically switched by pure electric currents. We consider $\text{FeS}$~\cite{Takagi2024NatMat}, CrSb~\cite{Reimers24NC,JianYD24prl}, $\text{RuO}_2$~\cite{Ahn19PRB,ShaoDF21NC,Libor22PRX2,ZiHL24arxiv}, MnF$_2$~\cite{yuanLD20PRB}, Rb$_{1-\delta}$V$_2$Te$_2$O~\cite{FaYZ24arxiv}, KV$_2$Se$_2$O~\cite{JiangB24arxiv}, $\text{RuF}_4$~\cite{Marko242DMat} and FeSb$_2$ \cite{Mazin21PNAS}, because they have large spin splitting and high N\'eel temperatures. The arrowed balls indicate the magnetic atoms and the cyan and pink shadows show the chemical environments created by surrounding non-magnetic atoms. Unfortunately, by far most predicted altermagnets have parity symmetry so that they cannot be switched by pure electric currents. The FeS bilayer, on the other hand, has the same symmetry as that of MnTe bilayer, so it may also be switchable, even at room temperature because of its high N\'eel temperature \cite{Takagi2024NatMat}. The CrSb bilayer, although has the broken parity symmetry, does not possesses anomalous Hall signal, due to its symmetries with a N\'eel order along the [0001] direction. }\label{Fig:candidates}
\end{figure*}

Moreover, we can analyze the switching results for currents applied along other directions, as shown in Fig.~\ref{Fig:Symmetry}(h) for currents in the $-x$ and $\pm y$ directions. The currents in the $\pm y$ direction switches the N\'eel order to the middle of two nearby stable positions (Sec.~S1 of \cite{supp}). After turning off the current, the N\'eel order will randomly relax to one of the two nearby stable positions. The two nearby stable positions, however, have opposite anomalous Hall signals \cite{Gonzalez2023PRL}, leading to failure of detectable 180$^\circ$ switching if the current is in the $\pm y$ directions (details in Sec.~S2 of \cite{supp}). Therefore, to achieve a deterministic 180$^\circ$ switching of altermagnetism that can be detected by the anomalous Hall effect in the MnTe bilayer, the current should be injected along the $x$ direction (or three  directions equivalent to the [11$\bar{2}$0] crystallographic direction).

{\color{blue}\emph{Magnetic dynamics simulation}} - To verify the switching of the $\pm y$-direction N\'eel orders driven by the $\pm x$-direction current in the MnTe bilayer, we simulate the dynamics of the switching, by using the Landau-Lifshitz-Gilbert equation \cite{Slonczewski1996JMMM,Wolf2001Sci}
\begin{eqnarray}\label{Eq:LLG}
(1+\alpha^2)\dot{\mathbf{m}}_{a}&=&\alpha \mathbf{m}_{a}\times\left(\frac{|\gamma|}{M_S} \mathbf{m}_{a}\times \frac{\delta H_m}{\delta\mathbf{m}_{a}}+\mathbf{T}_{a}\right)\nonumber \\
&&+\frac{|\gamma|}{M_S} \mathbf{m}_{a}\times \frac{\delta H_m}{\delta\mathbf{m}_{a}}+\mathbf{T}_{a},
\end{eqnarray}
where $\alpha=0.1$ is the Gilbert damping, $\gamma (<0)$ is the gyromagnetic ratio of the electron, $M_S=3.89\mu_B$ is the saturation moment of a Mn atom with $\mu_B$ the Bohr magneton, $H_m$ is the magnetic Hamiltonian that describes the magnetic structure of the MnTe bilayer
\begin{eqnarray}\label{Eq:H-m}
    H_m&=&J_m \mathbf{m}_A\cdot\mathbf{m}_B+K\left[(\mathbf{m}_A\cdot\hat{\mathbf{z}})^2+(\mathbf{m}_B\cdot\hat{\mathbf{z}})^2\right]\notag \\
    &&+\frac{\lambda}{2}\left[\cos(6\varphi_A)+\cos(6\varphi_B)\right].
\end{eqnarray}
The first term of Eq.~(\ref{Eq:H-m}) describes the antiferromagnetic exchange coupling between $\mathbf{m}_A$ and $\mathbf{m}_B$. The exchange energy $J_m\approx 4.46$ meV is calculated by using the Green’s function method implemented in the TB2J package \cite{tb2j}. The tight-binding Hamiltonian in the atomic basis, as delineated in Eq.~(\ref{Eq:TB}), is employed for the calculation. The second term with $K=0.1$ meV \cite{Kriegner17prb} describes the anisotropy that keeps the magnetic structure within $x$-$y$ plane. The third term with $\lambda=0.01$ meV \cite{Kriegner17prb} describes the six-fold symmetric anisotropy of magnetic structure within $x$-$y$ plane. $\mathbf{T}_{a}$ in Eq.~(\ref{Eq:LLG}) is the spin-orbit torques induced by the electric current
\begin{eqnarray}\label{SOT-H}
\mathbf{T}_{a}\equiv \frac{|\gamma|}{M_S} \mathbf{m}_{a}\times \frac{\delta E_c}{\delta\mathbf{m}_{a}},
\end{eqnarray}
where the energy generated by the current-induced spin accumulations per unit cell can be found as
\begin{eqnarray}\label{Eq:Ec}
    E_{\mathrm{c}}(\varphi)=(2JV_{\mathrm{U}}/ \hbar) \left[\mathbf{m}_A\cdot\mathbf{s}_A(\varphi)+\mathbf{m}_B\cdot \mathbf{s}_B(\varphi) \right],
\end{eqnarray}
where $V_\mathrm{U}$ is the volume of the unit cell, and the exchange energy $J$ between the Mn magnetic moments and spins of itinerary electrons is extracted from Eq.~(\ref{Eq:TB}). Specifically, the Fermi surface of MnTe is mainly composed of five $3d$ orbitals, so $J$ is evaluated by the coupling between the spin-up and spin-down states on the $3d$ orbitals, when the Mn atoms are polarized along the $x$ direction. It ranges from -1.830 eV to -1.691 eV and its average value is -1.756 eV. In this case the spin-orbit torques takes the form $\mathbf{T}_{a}=(2JV_\mathrm{U}|\gamma|/M_S\hbar) \mathbf{m}_{a}\times \mathbf{s}_a$ \cite{Haney2008JMMM,Manchon2019handbook,Manchon2019RMP}.

Figure \ref{Fig:Simulation}  shows that the N\'eel order can be switched from $\varphi=3\pi/2$ (equivalent to $-\pi/2$) to $\pi/2$ by a $+x$ direction current applied from 40 to 80 ps (cyan area) and from $\varphi=\pi/2$ to $-\pi/2$ by a $-x$ direction current from 120 to 160 ps (pink area). Moreover, near 50 ps and 130 ps, the N\'eel order oscillates to the stable positions, consistent with the symmetry analysis of the switching tendency in Fig.~\ref{Fig:Symmetry}(a)-(d). The switching can be achieved only when the current density is above a threshold. For $J$ ranges from -1.830 eV to -1.691 eV, the threshold current has a range
\begin{eqnarray}
j_c \approx 2.4 - 2.6 \times10^5\  \mathrm{A/cm}^2,
\end{eqnarray}
which can be understood by an energy analysis in \hyperlink{Appendix C}{Appendix C}.
Here, the threshold is lower than those in other antiferromagnetic devices \cite{Wadley2016Scn,Zheng2024NC,Higo2022Nature,Tsai2020Nature,Han2024SciAdv}, probably because we consider only single crystals with single magnetic domains. For realistic samples with polycrystals and multidomains, the threshold will be higher.

{\color{blue}\emph{Other altermagnet candidates}} - Recent studies have been suggesting an increasing number of candidate materials for altermagnets~\cite{YaQG23MTP,Mazin21PNAS,YaoYG24PRL,Osumi24PRB,Krempask24Nat,Lee24PRL,Moreno16PCCP,Marko242DMat,Reimers24NC,JianYD24prl,Naka20PRB,Das24PRL,JiangB24arxiv,FaYZ24arxiv,JiangB24arxiv,FaYZ24arxiv}, including $\text{FeS}$~\cite{Takagi2024NatMat}, CrSb~\cite{Reimers24NC,JianYD24prl}, $\text{RuO}_2$~\cite{Ahn19PRB,ShaoDF21NC,Libor22PRX2,ZiHL24arxiv}, MnF$_2$~\cite{yuanLD20PRB}, Rb$_{1-\delta}$V$_2$Te$_2$O~\cite{FaYZ24arxiv}, KV$_2$Se$_2$O~\cite{JiangB24arxiv}, $\text{RuF}_4$~\cite{Marko242DMat} and FeSb$_2$ \cite{Mazin21PNAS}. Most of the altermagnet candidates are parity symmetric and thus can not be deterministically switched by a pure current, as shown in Fig.~\ref{Fig:candidates}. To discover the altermagnetic materials that can be electrically switched, we should aim for the materials with broken parity symmetry, e.g., a MnTe bilayer as we study in this work. Nevertheless, parity symmetry breaking is merely a necessary requirement. For example, a CrSb bilayer has no parity symmetry, but the easy axis of its N\'eel order is in the crystallographic [0001] direction, as shown in Fig.~\ref{Fig:candidates}, leading to the absence of the anomalous Hall signal (details in Sec.~S5 of \cite{supp}). However, the Cr magnetic moments can be deviated from the [0001] direction, i.e., by the Dzyaloshinskii-Moriya interactions, giving a readable anomalous Hall signal \cite{Zhou25nat}.

\begin{acknowledgments}
We thank fruitful discussions with Qihang Liu, Chang Liu, Xiangang Wan, and Xiaoqun Wang. This work was supported by the National Key R\&D Program of China (2022YFA1403700), Innovation Program for Quantum Science and Technology (2021ZD0302400), the National Natural Science Foundation of China (12525401 and 12350402), Guangdong province (2020KCXTD001), Guangdong Basic and Applied Basic Research Foundation (2023B0303000011), Guangdong Provincial Quantum Science Strategic Initiative (GDZX2201001 and GDZX2401001), the Science, Technology and Innovation Commission of Shenzhen Municipality (ZDSYS20190902092905285), and the New Cornerstone Science Foundation through the XPLORER PRIZE. The numerical calculations were supported by Center for Computational Science and Engineering of SUSTech.

$^\dag$Yiyuan Chen and Xiaoxiong Liu contributed equally.
\end{acknowledgments}

\bibliographystyle{apsrev4-1-etal-title_10authors}
\bibliography{refs-RuO2}

%\newpage
\appendix

\section{\large \textbf{End Matter}}

\renewcommand{\theequation}{A\arabic{equation}}
\setcounter{equation}{0}

\hypertarget{Appendix A}{{\color{blue}\emph{Appendix A: Calculation of spin accumulations}}} - We calculate the current-induced spin accumulations on $a\in\{A,B\}$ Mn atom, by applying the first-principles calculations to the Kubo formula \cite{Mahan,Chen24prbl}
\begin{eqnarray}\label{Eq:Kubo-d}
(s^a_x,s^a_y,s^a_z)\equiv\mathbf{s}_a&=&\frac{e\hbar}{2V}\tau \sum_{\nu,\mathbf{k}}\frac{\partial f_\nu}{\partial\epsilon_\nu}(\mathbf{E}\cdot \mathbf{v}_{\nu}) \bm{\sigma}_{\nu}^{a},
\end{eqnarray}
where $e$ is the elementary charge, $\hbar$ is the reduced Planck constant, $V$ is the volume of the system, $\tau$ is the relaxation time, $f_\nu=1/\{\exp[(\epsilon_\nu-E_F)/k_BT]+1\}$ is the Fermi-Dirac distribution function, $\epsilon_\nu$ is the eigen energy of eigen state $|\nu\rangle$, the electric field $\mathbf{E}$ can be converted from the switching current density $\mathbf{j}=\sigma\mathbf{E}$, $\sigma$ is the conductivity, $\mathbf{v}$ is the group velocity operator, $\mathbf{v}_{\nu} = \langle \nu | \mathbf{v} |\nu\rangle $, $\bm{\sigma}^a =(\sigma^a_x, \sigma^a_y,\sigma^a_z)$ are the Pauli matrices for the local spin density on Mn atom $a$,
and $\bm{\sigma}^a_{\nu} = \langle \nu | \bm{\sigma}^a |\nu\rangle  $.

To evaluate the spin accumulations in the MnTe bilayer accurately, we employ the Wannier interpolation scheme~\cite{SpectralWannier,QiaoSHE} based on the \emph{ab initio} disentangled Wannier functions (details in Sec.~S3 of \cite{supp}).
The Wannier Hamiltonian reads
\begin{equation}\label{Eq:TB}
\mathcal{H}_{a\alpha , b \beta }^{\mathrm{W}}(\mathbf{R}) =\langle\mathbf{0} a \alpha | \hat{\mathcal{H}}|\mathbf{R} b \beta \rangle,
\end{equation}
describing the hoppings from localized orthonormal Wannier function $|\mathbf{R} b \beta \rangle$ ($\beta$ orbital of atom $b$ at lattice $\mathbf{R}$) to $|\mathbf{0} a \alpha \rangle$ ($\alpha$ orbital of atom $a$ at home unit cell $\mathbf{0}$). In momentum space, the eigen energies $\epsilon_\nu$ and eigen vectors $|\nu\rangle $ can be found from the eigen equation $\mathcal{H}{(\mathbf{k})} |\nu \rangle = \epsilon_\nu |\nu \rangle$, where the Fourier transform of the Wannier Hamiltonian
\begin{align}\label{Eq:H_k}
\mathcal{H}{(\mathbf{k})} = \sum_{\mathbf{R}} e^{i \mathbf{k} \cdot \mathbf{R}} \mathcal{H}_{a\alpha , b \beta }^{\mathrm{W}}(\mathbf{R}).
\end{align}

The electronic structures of MnTe slab with various directions of the magnetic moments are calculated
via density-functional theory in the pseudopotential framework and incorporating spin-orbit coupling.
In these calculations, we use the projected augmented wave method, implemented in the Vienna \textit{ab initio} simulation package~\cite{kresse1996efficiency,kresse1999ultrasoft,kresse19962}, adopting the Perdew-Burke-Ernzerhof parameterization of the generalized gradient approximation exchange-correlation functional~\cite{gga-pbe}. To account for electron-electron correlation on Mn sites, the Dudarev method is applied with a Hubbard effective parameter term $U = 3$~eV is applied on $d$ orbitals of Mn atoms.
The electron wave function is expanded on a 6$\times$6$\times$1 k-mesh, and the self-consistent computation is carried out until the energy difference is less than 10$^{-6}$~eV with a cut-off energy of 400~eV.

To freeze the magnetic moment of MnTe in a random direction, we first calculate the collinear spin-polarized scalar-relativistic electronic structure of the antiferromagnetic MnTe bilayer. The charge density in real space $\rho^0_{(\mathbf{r})}$ is a scalar on a sampling grid. Then, we project the scalar charge density $\rho^0_{(\mathbf{r})}$ to a vector charge density, following the direction of the magnetic moment,
\begin{align}\label{Eq:rho}
\rho^x_{(\mathbf{r})} &= \rho^0_{(\mathbf{r})} \sin{\theta} \cos{\phi},\nonumber\\
\rho^y_{(\mathbf{r})} &= \rho^0_{(\mathbf{r})} \sin{\theta} \sin{\phi},\nonumber\\
\rho^z_{(\mathbf{r})} &= \rho^0_{(\mathbf{r})} \cos{\theta},
\end{align}
where $\theta$ and $\phi$ are the polar and azimuthal angles of the magnetic moment, respectively. The $\rho^x_{(\mathbf{r})}$, $\rho^y_{(\mathbf{r})}$, and $\rho^z_{(\mathbf{r})}$ are well initial guess for the local spin density approximation (LSDA) method to calculate the non-collinear spin-polarized full-relativistic electronic structure of the MnTe bilayer with a random direction of the magnetic moment. The Wannier functions are disentangled from the Kohn-Sham wave functions, using $s,d$ orbitals of Mn and $p$ orbitals of Te as projections, by employing Wannier90 code package \cite{Wannier90}. 
To achieve accurate mapping of spin accumulations on the Fermi surface in the MnTe slab, we evaluate Eq.~(\ref{Eq:Kubo-d}) using Wannier interpolation method implemented in WannierBerrri~\cite{wannierberri}. In the calculation, a 200×200×1 k-point grid and 10 refinement iterations are used. The tetrahedron method~\cite{tetrahedronmethod,tetrahedronmethod2} is further applied to accurately describe the distribution function ${\partial f_\nu}/{\partial\epsilon_\nu}$ at zero temperature.
The scheme in Fig.~\ref{Fig:Symmetry} are from the diagonal spin accumulations. We show that the off-diagonal spin accumulations
 are three orders smaller and cannot support deterministic switching in Sec.~S3 and Sec.~S4 of \cite{supp}.

\renewcommand{\theequation}{B\arabic{equation}}
\setcounter{equation}{0}

\begin{figure}[htbp]
\includegraphics[width=0.4\textwidth]{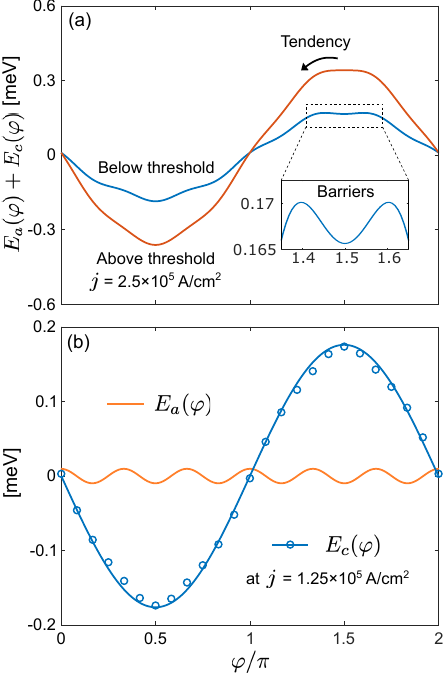}
\caption{(a) Total energy $E_c+E_a$ of the magnetic structure of MnTe in the presence of the switching current as a function of $\varphi$ for different current density, where $E_c$ is the energy of current-induced spin accumulations and $E_a$ is the anisotropy of the magnetic structure. The curved arrow indicates the switching tendency from the energy maximum at $\varphi=3\pi/2$. Inset: the local energy barriers near $\varphi=3\pi/2$ when the current density is below the threshold $j_c\approx2.5\times10^5$ A/cm$^2$.
(b) $E_c$ at
$j=$1.25$\times10^5$ A/cm$^2$ and $E_a$, which gives rise to the energy barrier.}\label{Fig:Threshold}
\end{figure}

\hypertarget{Appendix B}{{\color{blue}\emph{Appendix B: Symmetry operations $m_x$, $T$, and $Tm_x$ in Fig.~\ref{Fig:Symmetry}}}} - (i) By applying the mirror reflection $m_x$ with respect to the $y$-$z$ plane, the N\'eel order can be transformed from $\varphi$ in the first quadrant [Fig.~\ref{Fig:Symmetry}(a)] to $-\varphi$ in the fourth quadrant [Fig.~\ref{Fig:Symmetry}(b)] and the spin accumulations transform as
\begin{eqnarray}
m_x &: & \varphi \rightarrow -\varphi, \nonumber\\
&& s^{A,B}_x(\varphi)=-s^{A,B}_x(-\varphi), \nonumber\\ &&s^{A,B}_y(\varphi)=s^{A,B}_y(-\varphi),
\end{eqnarray}
as verified by the calculated results at the dashed line (b) in Figs.~\ref{Fig:Symmetry}(f) and \ref{Fig:Symmetry}(g). Therefore, the N\'eel order in the fourth quadrant also rotates anticlockwisely towards the $+y$ direction.

(ii) By applying the time reversal $T$, the N\'eel order can be transformed from $\varphi$ in the first quadrant [Fig.~\ref{Fig:Symmetry}(a)] to $\pi+\varphi$ in the third quadrant [Fig.~\ref{Fig:Symmetry}(c)] and the spin accumulations transform as
\begin{eqnarray}
T &: & \varphi \rightarrow \pi+\varphi, \nonumber\\
&& s^{A,B}_x(\varphi)=s^{A,B}_x(\pi+\varphi), \nonumber\\ &&s^{A,B}_y(\varphi)=s^{A,B}_y(\pi+\varphi),
\end{eqnarray}
as verified by the calculated results at the dashed line (c) in Figs.~\ref{Fig:Symmetry}(f) and \ref{Fig:Symmetry}(g). Therefore, the N\'eel order in the third quadrant also rotates clockwisely towards the $+y$ direction.

(iii) By applying $Tm_x$, the N\'eel order can be transformed from $\varphi$ in the first quadrant [Fig.~\ref{Fig:Symmetry}(a)] to $\pi-\varphi$ in the second quadrant [Fig.~\ref{Fig:Symmetry}(b)] and the spin accumulations transform as \begin{eqnarray}
Tm_x &: & \varphi \rightarrow \pi-\varphi, \nonumber\\
&& s^{A,B}_x(\varphi)=-s^{A,B}_x(\pi-\varphi), \nonumber\\ &&s^{A,B}_y(\varphi)=s^{A,B}_y(\pi-\varphi),
\end{eqnarray}
as verified by the calculated results at the dashed line (d) in Figs.~\ref{Fig:Symmetry}(f) and \ref{Fig:Symmetry}(g). Therefore, the N\'eel order in the second quadrant also rotates clockwisely towards the $+y$ direction.

\renewcommand{\theequation}{C\arabic{equation}}
\setcounter{equation}{0}

\hypertarget{Appendix C}{{\color{blue}\emph{Appendix C: Estimate of threshold current density}}} -
To understand the threshold current density in the magnetic dynamics simulation, we perform an energy analysis. The current-induced spin accumulations can give rise to an energy. When the energy of spin accumulations overcomes the local energy barrier produced by the in-plane anisotropy of the magnetic structure, the switching can be achieved.

The total energy of magnetic structure per unit cell in the presence of the switching current can be found as
\begin{eqnarray}
    E_{\mathrm{a}}(\varphi)+E_{\mathrm{c}}(\varphi),
\end{eqnarray}
where the anisotropy of the magnetic structure shows a six-fold rotational symmetry in the (0001) plane and can be described by
\begin{eqnarray}
E_{\mathrm{a}}(\varphi)=\lambda\cos(6\varphi),
\end{eqnarray}
where $\lambda=0.01$ meV \cite{Kriegner17prb}. The energy of the current-induced spin accumulations has been given in Eq.~(\ref{Eq:Ec}) and $J=-1.756$ eV.

For a $+x$ direction current, $E_{\mathrm{c}}(\varphi)+E_{\mathrm{a}}(\varphi)$ in Fig. \ref{Fig:Threshold}(a) shows a minimum at $\varphi=\pi/2$ and a maximum at $\varphi=3\pi/2$, so the switching tendency is from the maximum at $\varphi=3\pi/2$ to the minimum at $\varphi=\pi/2$. However,
below a threshold current density, there is a local energy barrier near $\varphi=3\pi/2$, as shown by the inset of Fig.~\ref{Fig:Threshold}(a). To flatten the barrier, the current density has to be above a threshold, then the switching can be achieved. Following this strategy, we numerically find that the threshold current density is about $j_c\approx2.5\times10^5\  \mathrm{A/cm}^2$. When $J\in$ [-1.830,-1.691] eV, $j_c\in[2.4,2.6]\times10^5\  \mathrm{A/cm}^2$.

\end{document}